\documentclass[11pt]{article}
\usepackage{multicol}
\usepackage{amssymb,amsmath,amsthm}
\usepackage{authblk}
\usepackage{multirow}
\usepackage{blindtext}
\usepackage{amsfonts}
\usepackage{amsmath,amssymb,xcolor}
\usepackage{slashed}
\usepackage{graphicx}

\newtheorem{theorem}{Theorem}

\newtheorem{corollary}[theorem]{Corollary}

\newtheorem{proposition}[theorem]{Proposition}
\newtheorem{remark}[theorem]{Remark}

\author{Frido Rolloos\thanks{frolloos@yahoo.com}}

\title{The ATM implied volatility slope, the (dual) volatility swap, and the (dual) zero vanna implied volatility}

\begin{document}
\maketitle

\abstract{Exact relationships between the short time-to-maturity ATM implied volatility slope, the (dual) volatility swap, and the (dual) zero vanna implied volatility are given.}

\section{Assumptions}
Various papers have studied the short time behavior of the at-the-money (ATM) implied volatility level and skew (see for example \cite{ALV} and \cite{MS}). Here we give another interpretation of the short-time ATM skew for diffusion models with stochastic volatility in terms of volatility swaps.

We will work under the premise that the market implied volatility surface is generated by the following general stochastic volatility (SV) model
\begin{equation}\label{model}
S_T = S_t \exp\left\{- \frac{1}{2} \int_t^T \sigma^2_u \, du + \int_t^T \sigma_u \left( \rho dW_u + \bar\rho dZ_u \right)\right\},
\end{equation}
where $dZ_u$ and $dW_u$ are standard Brownian motions, $\bar\rho = \sqrt{1-\rho^2}$. The instantaneous volatility $\sigma_u$ is assumed to be a positive process that is adapted to the filtration generated by $W_u$ and independent of the stock price $S_u$.

\smallskip
The SV process is assumed to be well-behaved in the sense that vanilla options prices are risk-neutral expectations of the payoff function:
\begin{equation}
C (S_t,t, K, T )= E_t \left[ \left( S_T - K \right)_+ \right]
\end{equation}
The option price $C$ can always be expressed in terms of the Black-Merton-Scholes (BS) price with an implied volatility parameter $I$:
\begin{equation}
C(S_t, K) = BS (S_t,K,I) 
\end{equation}
The BS price is given by the well-known formula
\begin{equation}
BS (S_t,K,I) = S_t  N(d_+) - K N(d_-),
\end{equation}
where $N \left(d_{\pm} \right)$ are normal distribution functions,
\begin{equation}
d_{\pm} = \frac{\log (S_t/K)}{ I \sqrt{T-t}} \pm \frac{I \sqrt{T-t}}{2}.
\end{equation}

\smallskip
In what follows we will use the following notations:
\begin{itemize}
\item The log spot price is defined as $x_t = \log S_t$ and the log strike as $k = \log K$.
\item The ATM implied volatility will be denoted as $I_0$ and the ATM log strike as $k_0$.
\item The zero vanna implied volatility $I_-$, the zero vanna log strike $k_-$, and their duals $I_+$ and $k_+$ are defined by
\begin{equation}
d_{\pm} (I_{\pm},k_{\pm}) = 0.
\end{equation}
\item The realised volatility over the interval $[t,T]$ is
\begin{equation}\label{rvol}
\sigma_{t,T} = \sqrt { \frac{1}{T-t}\int_t^T \sigma_u^2 \, du}.
\end{equation}
\item The volatility swap price is $E_t \left[ \sigma_{t,T} \right]$. The dual volatility swap price is $E_t \left[e^{x_T - x_t} \sigma_{t,T} \right]$. Notice that the dual volatility swap is the volatility swap under the share measure.
\end{itemize}

\section{Results}

\subsection{Approximations}
As derived using straightforward Taylor expansion in \cite{R} there is a simple relationship between (dual) volatility swap prices and (dual) zero vanna implied volatilities:
\begin{equation}
E_t \left[ \sigma_{t,T} \right] \approx I_- , \quad E_t \left[e^{x_T - x_t} \sigma_{t,T} \right] \approx I_+.
\end{equation}
Furthermore,
\begin{equation}
I_+ - I_- \approx \rho \, E_t \left[  \sigma_{t,T} \int_t^T \sigma_u \, dW_u \right].
\end{equation}

\smallskip
It is reasonable to assume that as $T - t \rightarrow 0$ the implied volatilities $I_-$ and $I_+$ are almost equidistant from the ATM implied volatility $I_0$. This is expressed as 
\begin{equation}\label{first}
I_- \approx I_0 - \frac{ \partial I_0}{\partial k} dk, \quad I_+ \approx I_0 + \frac{ \partial I_0}{\partial k} dk,
\end{equation}
and thus
\begin{equation}\label{second}
I_+ - I_- \approx 2 \frac{ \partial I_0}{\partial k} dk.
\end{equation}
From the definition of $I_{\pm}$ it follows that
\begin{align}\
x_t - (k_0 - dk) &= \frac{1}{2} I_-^2 (T-t), \\
x_t - (k_0 + dk) &= -\frac{1}{2} I_+^2 (T-t).
\end{align}
Inserting \eqref{first} into the above and ignoring terms of order $(dk)^2$ we obtain
\begin{equation}
2 dk \approx I_0^2 (T-t).
\end{equation}
Hence, by making use of \eqref{second}
\begin{equation}\label{res1}
I_0^2 (T-t) \,\frac{ \partial I_0 }{\partial k} \approx I_+ - I_-
\end{equation}
Remembering that $I_0 \rightarrow \sigma_{t} $ as $T-t \rightarrow 0$, we can also write
\begin{equation}\label{res2}
\frac{ \partial I_0 }{\partial k} \approx \frac{\rho}{ \sigma_t^2 (T-t)}\, E_t \left[  \sigma_{t,T} \int_t^T \sigma_u \, dW_u \right].
\end{equation}

\subsubsection{Example: lognormal instantaneous variance}
Suppose the instantaneous variance is lognormal,
\begin{equation}
d\sigma^2_u = \alpha \sigma^2_u \, dW_u.
\end{equation}
According to the integration by parts formula
\begin{equation}
E_t \left[  \sigma_{t,T} \int_t^T \sigma_u \, dW_u \right] = E_t \left[  \int_t^T \left( D^W_u \sigma_{t,T} \right) \sigma_u \, du, \right]
\end{equation}
with $D^W_u$ the Malliavin derivative with respect to $W_u$. Now
\begin{equation}
D^W_u \sigma_{t,T}  = \frac{1}{2 \sigma_{t,T} (T-t)} \int_u^T D^W_u \sigma_r^2 \, dr.
\end{equation}
For the the lognormal instantaneous variance model $D^W_u \sigma_r^2  = \alpha \sigma_r^2$, which for $T-t \rightarrow 0$ we can approximate by $\alpha \sigma_t^2$. Similarly we can approximate $\sigma_{t,T} \approx \sigma_t$. Hence,
\begin{equation}
E_t \left[  \sigma_{t,T} \int_t^T \sigma_u \, dW_u \right] \approx \frac{1}{4} \alpha \sigma^2_t (T-t).
\end{equation}
By making use of \eqref{res2} the short-time ATM implied volatility slope is
\begin{equation}
\frac{ \partial I_0}{\partial k} \approx \frac{\rho \alpha}{4}.
\end{equation}

\subsection{Exact results}

Assuming certain technical conditions are satisfied by the instantaneous volatility process, \cite{ALV} prove the following result on the short time to maturity behaviour of the slope of the ATM implied volatility.
\begin{theorem}\label{thmAlos}
Consider the model \eqref{model}. Then 
\begin{equation}
\lim_{T\downarrow t} \, \frac{1}{(T-t)^{H - \frac{1}{2}}} \, \frac{ \partial I_0}{ \partial k} = \frac{\rho}{2 \sigma_t^2} \lim_{T \downarrow t} \, \frac{1}{(T-t)^{H + \frac{3}{2}}} \, E_t \left[ \int_t^T \int_s^T D^W_s \sigma^2_r \, dr ds \right],
\end{equation}
where $H$ is the Hurst parameter and $D^W_u$ is the Malliavin derivative with respect to $W_u$.
\end{theorem}
\noindent
Restricted to the case $H=1/2$, Theorem~\ref{thmAlos} gives
\begin{equation}
\lim_{T\downarrow t} \,\frac{ \partial I_0}{ \partial k} = \frac{\rho}{2 \sigma_t^2} \lim_{T \downarrow t} \, \frac{1}{(T-t)^2} \, E_t \left[ \int_t^T \int_s^T D^W_s \sigma^2_r \, dr ds \right].
\end{equation}

\smallskip
We now derive exact relationships between the slope of the ATM implied volatility, volatility swaps under the money market and share measures, and the zero vanna implied volatility and its dual for the case $H=1/2$. The same hypotheses on the instantaneous volatility process as in \cite{ALV} are assumed to hold.
\begin{proposition}\label{p1}
\begin{equation}
\lim_{T \downarrow t} \frac{E_t \left[ \left( e^{x_T - x_t} - 1 \right) \sigma_{t,T}\right]}{\sigma_t^2 (T-t)} = \lim_{T\downarrow t} \, \frac{ \partial I_0}{ \partial k}.
\end{equation}
\begin{proof}
Notice that we can write the model \eqref{model} as
$$
x_T = \widetilde x_T + M_T,
$$
with
$$
\widetilde x_T = x_t - \frac{\bar\rho^2}{2} \int_t^T \sigma^2_u \, du + \bar\rho  \int_t^T \sigma_u \,  dZ_u ,
$$
and
$$
M_T = - \frac{\rho^2}{2} \int_t^T \sigma^2_u \, du + \rho  \int_t^T \sigma_u \,  dW_u.
$$
Thus,
\begin{align*}
E_t \left[ e^{x_T - x_t} \sigma_{t,T} \right] &= E_t \left[ e^{\widetilde x_T - x_t}e^{M_T} \sigma_{t,T}\right] \\
&= E_t \left[ e^{M_T} \sigma_{t,T}\right].
\end{align*}
where the second equality is obtained by conditioning on $\mathcal{F}_T^W$. Since 
$$
e^{M_T} = 1 + \rho \int_t^T \sigma_u e^{M_u} \, dW_u,
$$
it follows that
$$
E_t \left[ \left( e^{x_T - x_t} - 1 \right)\sigma_{t,T}\right] = \rho E_t \left[  \sigma_{t,T} \int_t^T \sigma_u e^{M_u} \, dW_u \right].
$$
A straightforward application of the integration by parts formula then gives us
$$
E_t \left[ \left( e^{x_T - x_t} - 1 \right) \sigma_{t,T} \right] = \frac{\rho}{2 (T-t)} E_t \left[ \frac{1}{\sigma_{t,T}} \int_t^T \left( \int_s^T D^W_s \sigma^2_r \, dr \right) e^{M_s} \sigma_s ds \right].
$$
Comparing this with Theorem~\ref{thmAlos} it follows that
\begin{align*}
\lim_{T \downarrow t} \frac{E_t \left[ \left( e^{x_T - x_t} - 1 \right) \sigma_{t,T} \right]}{\sigma_t^2 (T-t)} &= \frac{\rho}{2 \sigma_t^2} \lim_{T \downarrow t} \, \frac{1}{(T-t)^2} \, E_t \left[ \int_t^T \int_s^T D^W_s \sigma^2_r \, dr ds \right] \\
&= \lim_{T\downarrow t} \, \frac{ \partial I_0}{ \partial k}.
\end{align*}
\end{proof}
\end{proposition}

\begin{remark}
The quantity $E_t \left[ \left( e^{x_T - x_t} - 1 \right) \sigma_{t,T}\right]$ is the covariance between the simple return of the spot price and its realised volatility. Proposition~\ref{p1} therefore says two things: First of all the slope of the ATM implied volatility contains information about the covariance between the spot and its volatility. Second, although the volatility swap and its dual are model dependent, their normalised difference becomes increasingly model independent as time to maturity decreases.
\end{remark}

\begin{proposition}
If the implied volatility skew admits a Taylor series in strike, then
\begin{equation}
\lim_{T \downarrow t} \, \frac{I_+ - I_-}{\sigma_t^2 (T-t)} =  \lim_{T\downarrow t} \, \frac{ \partial I_0}{ \partial k}.
\end{equation}
\begin{proof}
Without loss of generality we can set $x_t = k_0 =0$. Write
$$
I_{\pm} = I_0 + k_{\pm}\frac{\partial I_0}{\partial k} + O\left( k_{\pm}^2 \right).
$$
From the definitions of $I_{\pm}$ and $k_{\pm}$ it follows that
\begin{align*}
I_+ - I_-  &= \frac12 (I_+^2 + I_-^2)(T-t) \frac{\partial I_0}{\partial k} +  O \left( (T-t)^2 \right) \\
&= I_0^2 (T-t) \frac{\partial I_0}{\partial k} + O \left( (T-t)^2 \right).
\end{align*}
Since $\lim_{T\downarrow t}\, I_0^2 = \sigma_t^2$ and $\lim_{T\downarrow t} \, \frac{ \partial I_0}{ \partial k}$ exists, 
$$
\lim_{T \downarrow t} \, \frac{I_+ - I_-}{T-t} = \sigma_t^2 \lim_{T\downarrow t} \, \frac{ \partial I_0}{ \partial k}.
$$
\end{proof}
\end{proposition}

\begin{corollary}
\begin{equation}
\lim_{T \downarrow t}\, \frac{ E_t \left[ \left( e^{x_T - x_t} - 1 \right) \sigma_{t,T}\right] }{T-t}= \lim_{T \downarrow t} \, \frac{I_+ - I_- }{T-t}.
\end{equation}
\end{corollary}

%\bibliography{mybib}

\begin{thebibliography}{11}

\bibitem{ALV} Al\`{o}s, E., Le\'{o}n, J. A., and Vives, J. ``On the short-time behavior of the implied volatility for jump-diffusion models with stochastic volatility,'' Finance and Stochastics, 2007

\bibitem{MS} Medvedev, A., and Scaillet, O. ``Approximation and calibration of short-term implied volatilities under jump-diffusion stochastic volatility," Review of Financial Studies, 2007

\bibitem{R} Rolloos, F. ``Managing forward volatility and skew risk,'' Available at SSRN, 2021

%\bibitem{RT} Romano, M., and Touzi, N. ``Contingent claims and market completeness in a stochastic volatility model,'' Math. Financ. 7, (1997): 399-410.

%\bibitem{W} Willard, G.A. ``Calculating prices and sensitivities for path-independent securities in multifactor models,'' J. Deriv. 5, (1997): 45-61.

\end{thebibliography}
%\bibliographystyle{abbrvnat}

\end{document}